\documentclass[12pt,preprint]{aastex}

\usepackage{emulateapj5}
\usepackage{psfig}

\newcommand{\psr}{PSR J0218$+$4232 }
\newcommand{\gtap}{\mathrel{\hbox{\rlap{\lower.55ex \hbox {$\sim$}}
                   \kern-.3em \raise.4ex \hbox{$>$}}}}
\newcommand{\ltap}{\mathrel{\hbox{\rlap{\lower.55ex \hbox {$\sim$}}
                   \kern-.3em \raise.4ex \hbox{$<$}}}}

\slugcomment{}

\shorttitle{Chandra \psr observations}
\shortauthors{Kuiper et al.}

\begin{document}



\title{High resolution spatial and timing observations of\\
       millisecond pulsar \psr with Chandra}


\author{L. Kuiper and W. Hermsen}
\affil{SRON-National Institute for Space Research, Sorbonnelaan 2, 3584
CA, Utrecht, The Netherlands}
\author{F. Verbunt}
\affil{Astronomical Institute Utrecht University, P.O.Box 80000, 3508
TA, 
       Utrecht, The Netherlands}
\and
\author{S. Ord\altaffilmark{1}, I. Stairs\altaffilmark{2} and A. Lyne}
\affil{University of Manchester, Jodrell Bank, Macclesfield SK11 9DL,
United
       Kingdom}

\altaffiltext{1}{present address: Centre for Astrophysics and
Supercomputing, 
Swinburne University of Technology, P.O. Box 218, Hawthorn, Victoria
3122, Australia}
\altaffiltext{2}{present address: NRAO, P.O.Box 2, Green Bank, WV
24944, U.S.A.}




\begin{abstract}
We report on high-resolution spatial and timing results for binary
millisecond
pulsar PSR J0218$+$4232 obtained with the Chandra HRC-I and HRC-S in
imaging mode. 
The sub-arcsecond resolution image of the HRC-I (0.08-10 keV) showed
that the 
X-ray emission from \psr is consistent with that of a point source
excluding
the presence of a compact nebula with a size of about $14\arcsec$ for
which we 
had indications in ROSAT HRI data. The presence of a DC component is
confirmed. 
This X-ray DC component has a softer spectrum than the pulsed emission
and
can be explained by emission from a heated polar cap.
With our HRC-S observation we obtained a 0.08-10 keV pulse profile with
high statistics 
showing the well-known double peaked morphology in more detail than
before: The two
pulses have broad wings and pulsed emission appears only to be absent
in a narrow phase 
window of width $\ltap 0.1$. 
The absolute timing accuracy of $\sim 200 \mu$s makes it possible to
compare
for the first time in absolute phase the X-ray pulse profile with the
highly structured 
radio profile and the high-energy $\gamma$-ray profile (0.1-1 GeV). The
two X-ray pulses 
are aligned within absolute timing uncertainties 
with two of the three radio pulses. Furthermore, the two $\gamma$-ray
pulses are aligned 
with the two non-thermal X-ray pulses, corresponding to a probability
for a random occurrence
of 4.9$\sigma$, strengthening the credibility of the earlier reported
first 
detection of pulsed high-energy $\gamma$-ray emission from a (this)
millisecond pulsar. 
\end{abstract}








\section{Introduction}

\begin{figure*}[t]
  \hbox{\hspace{0.2cm}{\psfig{file=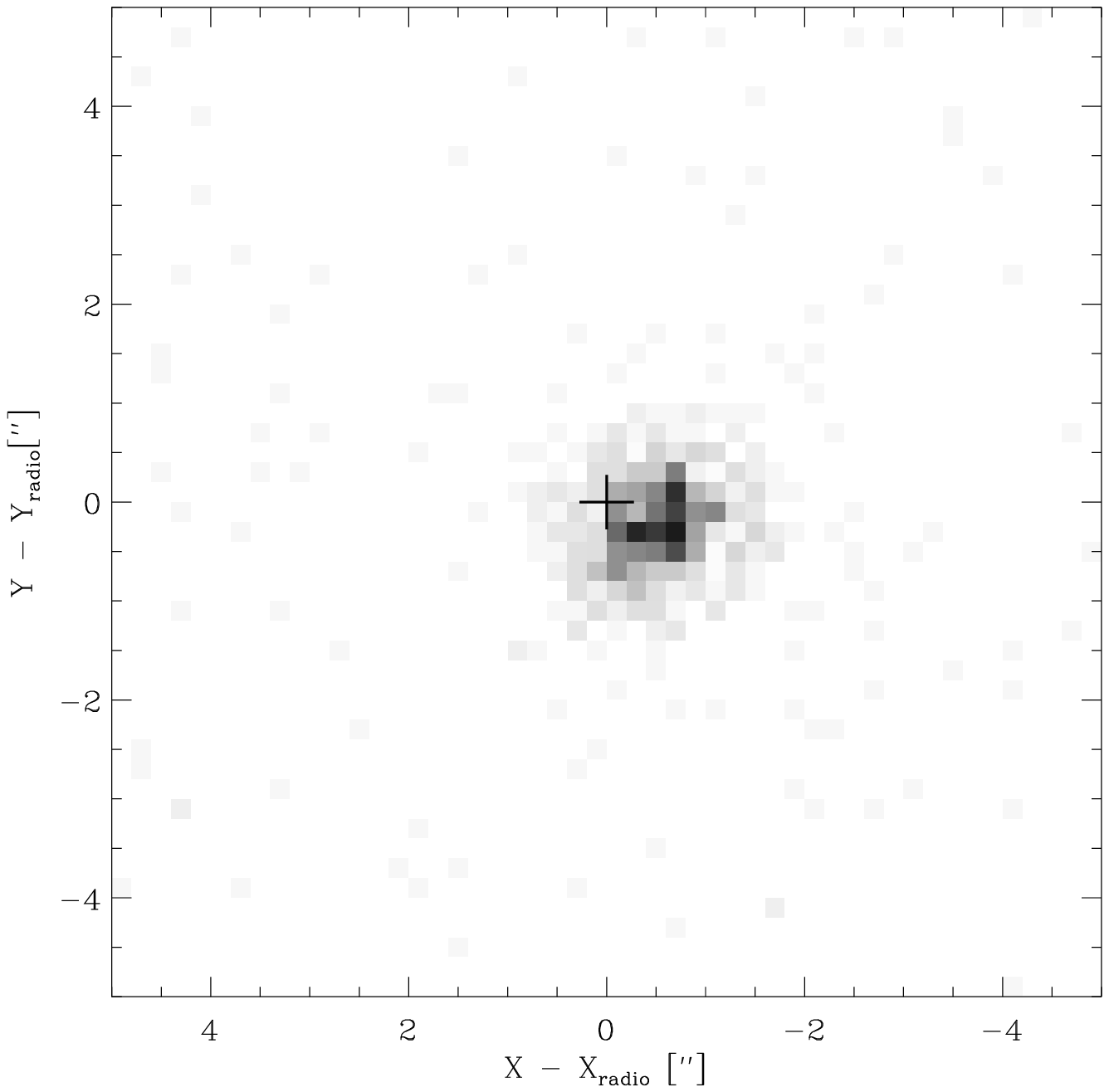,width=8.0cm,height=8.0cm}} 
          \hspace{1.00cm}
         {\psfig{file=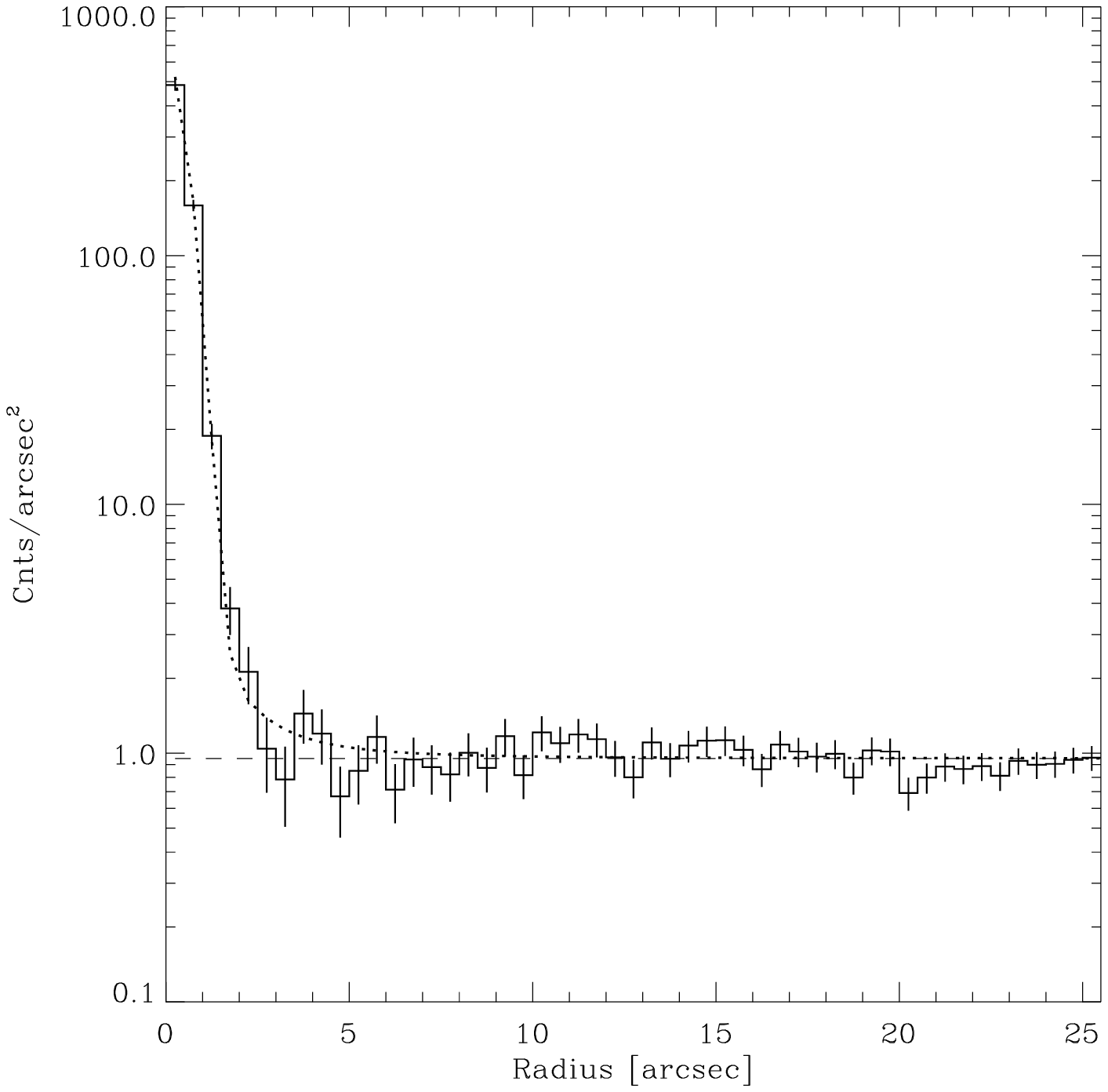,width=8.0cm,height=7.5cm}}
       }

  \caption{(left) Chandra 0.08-10 keV HRC-I image of a $10\arcsec\times 10\arcsec$ 
           region centered on the radio pulsar position of \psr. The radio position is marked
           with a `+' sign. The angular distance between the radio pulsar position and the X-ray 
           centroid is $\sim 0\farcs 6$, consistent with the Chandra localization accuracy. 
           (right) Radial distribution of HRC-I (Level 1) events using the best fit maximum 
           likelihood X-ray position as centre. Superposed as dotted line is 
           the radial profile from the PSF determined from the 2D fit procedure, which is consistent
           with the Chandra response to a point source.  The dashed line indicates the background 
           level derived from counts in the range 10 -- 25 arcsec from the centre. \label{hrcispatial}
          }
\end{figure*}


\psr is a 2.3 ms pulsar in a two day orbit around a low mass ($\sim 0.2$ M$_{\sun}$) 
white dwarf companion \citep{nav1995,kerk1997}. Pulsed X-ray emission with a 
Crab-like double pulse profile has been reported from ROSAT 0.1-2.4 keV data 
\citep{kuip1998a} and BeppoSAX MECS 1.6-10 keV data \citep{min2000}. The pulsed 
spectrum as measured by the MECS appeared to be remarkably hard with a power-law 
photon index $0.61 \pm 0.32$, harder than measured for any other radio pulsar.
Furthermore, \citet{kuip2000} report the detection with EGRET of pulsed
high-energy (0.1-1 GeV) $\gamma$-ray emission from this ms pulsar.
Their argument is based on three lines of evidence: (1) the 0.1-1 GeV data show a
3.5$\sigma$ pulsation at the radio period; (2) the $\gamma$-ray light curve
resembles the one seen in hard X-rays, namely a phase separation of $\sim$ 0.45 
between two pulses/maxima; (3) the spatial analysis shows that the position of the 
EGRET source 3EG J0222+4253 moves from the position of the nearby BL Lac 3C 66A 
towards the pulsar position with decreasing gamma-ray energy (for energies between 
100 and 300 MeV all source counts could be attributed to PSR J0218+4232).
They also showed that the two $\gamma$-ray pulses/maxima appeared to be aligned in 
absolute phase with two of the three radio pulses detected at 610 MHz.
Unfortunately, the timing accuracies of the ROSAT and BeppoSAX observations were
insufficient to construct X-ray profiles in absolute phase. 
None of the current models for pulsed X-ray and $\gamma$-ray emission from radio 
pulsars offers a consistent explanation for the above summarized high-energy results on
 \psr \citep{kuip2000}.

\psr is also remarkable in that it is the only Crab-like ms pulsar with a large
DC (unpulsed) fraction of $63 \pm 13\%$ in the ROSAT band below 2.4 keV
\citep{kuip1998a}, as well as a large DC fraction of $\sim 50\%$ in radio, systematically
over the range 100-1400 MHz \citep{nav1995}. 
The DC components as measured in the ROSAT and radio observations could be explained by 
emission from a compact nebula with diameter $\sim 14\arcsec$, but in both cases the
indications were at the limit of the imaging capabilities. 
Assuming that the radio DC component is compact, combined with the measured very broad 
and structured radio pulse profile, \citet{nav1995} suggested that the magnetic 
field of \psr is almost aligned with the rotation axis, the observer viewing the system 
under a small angle with respect to the rotation axis.
\citet{stairs1999} measured the magnetic inclination angle analyzing radio polarization 
profiles. Their rotation vector model fits indicate that the magnetic
inclination angle is indeed consistent with $0\arcdeg$ ($8\arcdeg\pm11\arcdeg$). 
Unfortunately, in their fits the line-of-sight inclination is unconstrained. If the DC 
component in X-rays is also compact, for the suggested geometry of a nearly aligned rotator 
and a small viewing angle, it can originate in the pulsar magnetosphere as well as from a 
heated polar cap of the neutron star. 

The objectives of our Chandra observations were: 1) To establish the spatial extent of 
the X-ray DC component, compact or extended; 2) To construct an X-ray pulse profile which can be 
compared in absolute phase with radio profiles and the 3.5$\sigma$ EGRET high-energy (0.1-1 GeV)
$\gamma$-ray profile.

\section{Observations}

\begin{figure*}
  \hbox{\hspace{0.5cm} \psfig{figure=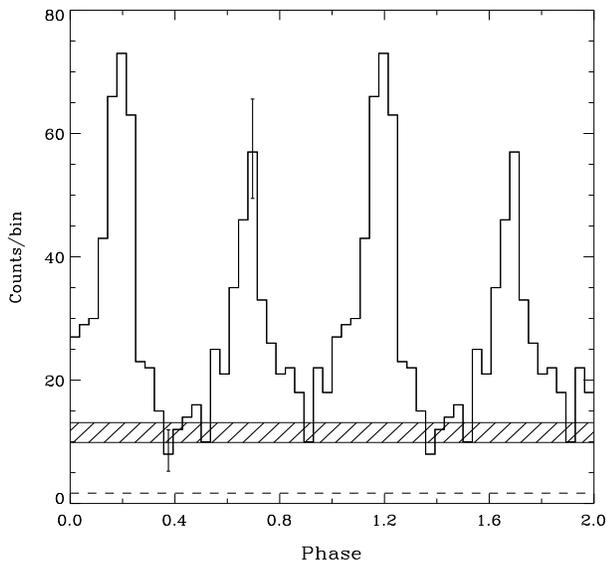,width=8.7cm,height=8.2cm}
  \hspace{0.3cm}{\parbox{80mm}{\vspace{-7cm}
  \caption{Pulse-profile of \psr as measured by the HRC-S in imaging
           mode. Two cycles are shown for clarity. Typical error bars 
           are shown at phases 0.35 and 0.7.
           The DC-level ($\pm 1\sigma$) is indicated by the hatched area, and the
           background level from the imaging analysis 
           by the dashed line.\label{hrcspulseprof}}}}}
\end{figure*}


\psr was observed with the HRC-I and HRC-S instruments on the Chandra
X-ray observatory (CXO) during two observations
of $\sim 75$ ks duration each. The HRCs are multi-channel plate (MCP)
detectors sensitive to X-rays 
in the 0.08-10 keV energy range with no spectral information.

The first observation with the HRC-I camera at the focal plane of 
the telescope mirror took place on 1999 December 22 for an effective
exposure of 74.11 ks.
Unfortunately, this observation suffered from a non-recoverable timing
(``wiring'') problem 
assigning incorrectly the event trigger time to that of the next
trigger \citep{ten2001}. 
On-board event screening prevents recovering the true event times of
all the triggers in 
the telemetry stream by simple 
back-shifting. The timing accuracy in this case is determined by the
total event rate, typically 
300-400 s$^{-1}$, and is much worse than the $16\mu$s intrinsic timing
resolution. This degradation
prevents the construction of high-resolution pulse profiles in case of
millisecond pulsars.  

In a re-observation of \psr on 2000 October 5 for an effective exposure
of 73.21 ks with the HRC-S 
in imaging mode (using only the central MCP segment) all event triggers
could be put in the 
telemetry stream, so that back-shifting recovers the intrinsic timing
accuracy. 

\section{Spatial Analysis with HRC-I}

\psr has clearly been detected near the centre of the $30\arcmin\times
30\arcmin$ field of view in
the 74.11 ks HRC-I observation. Within $14\farcm 5$ from the pulsar 51
X-ray sources have been 
detected, some of them have counterparts in our previous ROSAT HRI
(0.1-2.4 keV) and 
BeppoSAX MECS (1.6-10 keV) observations \citep{kuip1998a,min2000}. The
central $\sim 7\arcmin\times 7\arcmin$ part of the 
X-ray image containing \psr has also been observed at optical
wavelengths by the Keck telescope \citep{kerk2001} 
and it turns out that all central sources except one have optical
counterparts. Relative astrometry demonstrates 
that the Chandra positions are accurate to within $0\farcs 5(1)$,
better than the celestial localization accuracy
determination requirement of $1\arcsec$. Using the standard Level 1
event file provided by the CXC pipeline analysis (this file 
contains all HRC triggers with the position information corrected for
instrumental (de-gap) and aspect (dither) 
effects), a zoom-in at the pulsar location (Fig. \ref{hrcispatial}
left) shows that the X-ray centroid has a 
$\sim 0\farcs 6$ offset from the radio-pulsar position of $2^{\mathrm
h}18^{\mathrm m}6\fs 351$, $42\arcdeg 
32\arcmin 17\farcs 45$ (epoch J2000) consistent with our previous
findings. 
It is also evident from the image shown in Fig. \ref{hrcispatial}
(left) that the source 
events are concentrated within a circle with a radius of $\sim
2\arcsec$ from the X-ray centroid. A more quantitative 
estimate involves maximum likelihood fitting of the measured 2D spatial
distribution with a model composed
of a Gaussian and a generalized Lorentzian with free shape parameters,
normalization and centroid 
position on top of an (assumed) flat background, also with free scale
factor. The optimum X-ray centroid position is 
$2^{\mathrm h}18^{\mathrm m}6\fs 305(1)$,$42\arcdeg 32\arcmin 17\farcs
23(2)$ (epoch J2000) while the radial profile 
of the best model fit is compatible with the PSF of the HRMA/HRC-I
combination (95\% of the source counts are 
within $2\arcsec$ from the X-ray centroid). Fig. \ref{hrcispatial}
(right) shows the best model profile (dotted line)
superposed on the measured radial profile using the optimum X-ray
centroid position as centre. Thus, we have 
no evidence for extended emission near \psr at $\sim 1\arcsec$ scales
(diameter), rejecting the indication for a
compact nebula found in our analysis of ROSAT HRI data
\citep{kuip1998a}.

The total number of counts assigned to the pulsar from the maximum
likelihood fit is $870\pm 30$, which 
translates in a count rate of $(1.175 \pm 0.041)\times 10^{-2}$.
Assuming absorption in a column with density
$\hbox{\rm N}_{\mathrm H}=5\cdot 10^{20}$ cm $^{-2}$ \citep{verb1996}
and a photon power-law index of $0.94$
\citep{min2000} the count rate converts to a 2-10 keV flux of 
$(5.2\pm 0.2\  ^{+2.1} _{-1.3})\times 10^{-13}$ erg cm$^{-2}$s$^{-1}$.
The first error represents the statistical error and the second the
systematic error due to uncertainties in
the absolute sensitivity of the HRC-I and in the spectral model. The
value is consistent with the BeppoSAX MECS value 
of $(4.38\pm 0.48\ \pm 0.44)\times 10^{-13}$ erg cm$^{-2}$s$^{-1}$ (the
first error represents the 
statistical uncertainty and the second error the systematic uncertainty
of about 10\%).

\section{Timing Analysis with HRC-S}

In a spatial analysis using the HRC-S Level 1 event file the X-ray
centroid of the X-ray counterpart of \psr 
was found to have an offset of $1\farcs 9$ from the radio position of
the pulsar, larger than the celestial
location accuracy determination requirement of $1\arcsec$; a systematic
difference noticed earlier by e.g. 
\citet{wang2001}. The number of counts assigned to the source, applying
a similar
likelihood procedure as in the case of the HRC-I image data, is $755 \pm
28$. Using the optimized background scale 
factor and source profile, the optimum radius for event extraction (S/N
optimal) from the best fit
centroid turns out to be $1\farcs 5$.

The first step in the timing analysis is to correct the assigned event
times by back-shifting, recovering the 
intrinsic relative time resolution of $16\mu$s. Next, the event
extraction radius was set to the optimum radius of 
$1\farcs 5$. The final step is the determination of the arrival times
at the solar system barycentre using the 
orbital information of Chandra and the position of PSR J0218$+$4232.
Folding the barycentered arrival times with the spin and binary
parameters from an updated ephemeris for \psr
revealed the well-known doubled peaked profile at high statistics
(Fig.\ref{hrcspulseprof}). The deviation from a 
flat distribution is $15.2\sigma$ according to a $Z^{2}_{6}$ - test and
the peak separation is $0.475\pm0.015$ 
consistent with previous estimates \citep{kuip1998a,min2000}. The peak
widths $\Delta\phi$ (FWHM), fitting
the profile in terms of two asymmetric Lorentzians atop a flat
background, are: Peak-1 (at $\phi\simeq 0.2$) 
$0.116\pm0.019$ ($267\pm44\mu$s) and Peak-2 (at $\phi\simeq 0.7$)
$0.109\pm 0.025$ ($251\pm58\mu$s). 
Fitting two Gaussians plus background yielded similar results.

\begin{figure*}
  \hbox{\hspace{1.0cm} \psfig{figure=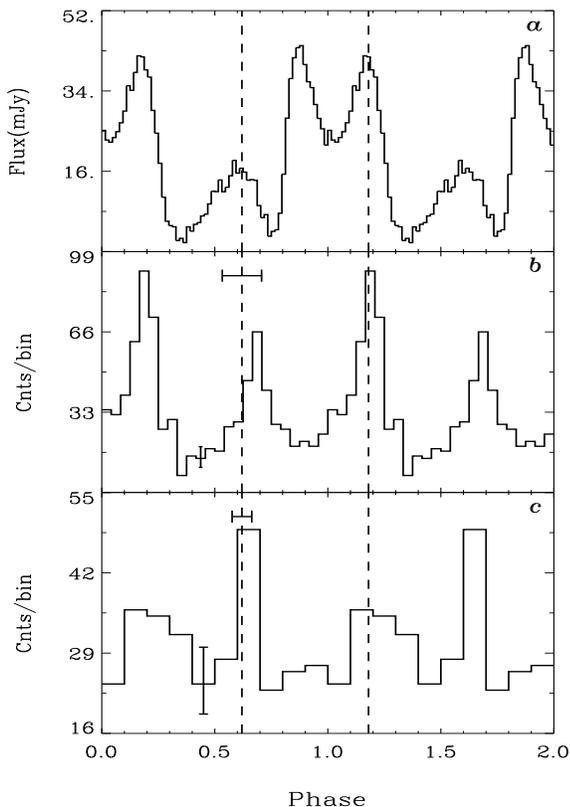,width=7.5cm,height=12cm}
  \hspace{1.1cm}{\parbox{80mm}{\vspace{-10cm}
  \caption{Multi-wavelength pulse profiles of \psr in absolute phase. 
         {\bf{a}}) Radio pulse profile at 610 MHz. {\bf{b}}) Chandra HRC-S 
          X-ray pulse profile (0.08-10 keV; timing accuracy $200\mu$s or 
          $0.085$ in phase) and {\bf{c}}) EGRET $\gamma$-ray pulse profile 
          (0.1-1 GeV; timing accuracy $100\mu$s or $0.043$ in phase).
          The absolute timing accuracies of the X- and $\gamma$-ray profiles 
          are shown as horizontal bars centered on phase 0.62. 
          Indicated as dotted lines are the positions of the 2 pulses in 
          the 610 MHz radio profile which coincide with the high-energy pulses. 
          Typical $\pm 1\sigma$ error bars are indicated in the X- and
          $\gamma$-ray profiles.  Note that the background level in the $\gamma$-ray profile
          as determined from an imaging analysis is at $22.2 $ \citep{kuip2000}.
          \label{hepulsprofstack}}}}}
\end{figure*}

%

In Fig.\ref{hrcspulseprof} is indicated an estimate for the unpulsed
(DC) level ($\pm 1\sigma$) using a bootstrap 
method outlined by \citet{swan1996}. Furthermore, the background level
estimated in the maximum likelihood 
imaging analysis is also shown, giving a number of background counts
within the $1\farcs 5$ extraction 
radius of $(46.9\pm0.4)$. The pointlike (see Sect. 3) DC (unpulsed)
component, significantly detected with ROSAT 
(0.1-2.4 keV; $4.8 \sigma$) by \citet{kuip1998a} and weakly with
BeppoSAX (1.6-4.0 keV; $\sim 2 \sigma$) by 
\citet{min2000} is clearly seen in this 0.08-10 keV profile. The pulsed
fraction $\cal{F}_P$ defined as 
${\cal{N}_P}/({\cal{N}_P+\cal{N}_{DC}})$, where $\cal{N}_P$ specifies
the number of pulsed source counts and $\cal{N}_{DC}$
the number of DC source counts, turns out to be ${\cal{F}_P} =
0.64\pm0.06$ as measured by CXO HRC-S over the entire 0.08-10 keV 
energy range.

\begin{table*}[h]
\caption{\label{ephemeris} Ephemeris of \psr based on radio timing measurements
spread over a 6.5 year period}
\begin{flushleft}
\begin{tabular}{ll}
\hline\noalign{\smallskip}
\noalign{\smallskip}
Parameter &  Value  \\
\hline
Right Ascension (J2000)        &  02$^{\rm h}$ 18$^{\rm m}$ 6\fs351    \\
Declination (J2000)            &  42$^\circ$ $32'$ 17\farcs45          \\
Epoch validity start/end (MJD) &  49092 -- 51462                       \\
Frequency                      &  430.4610670731491 Hz                 \\
Frequency derivative           &  $-1.43397\times 10^{-14}$ Hz s$^{-1}$\\
Epoch of the period (MJD)      &  50277.000000025                      \\
Orbital period                 &  175292.302218 s                      \\
a $\cdot$ sin i                &  1.9844301 (lt-s)                     \\
Eccentricity                   &  0                                    \\
Longitude of periastron        &  0                                    \\
Time of ascending node (MJD)   &  50276.61839862                       \\
RMS of timing solution         &  14.4 milliperiods; 33.4 $\mu$s       \\
\noalign{\smallskip}
\hline
\noalign{\smallskip}
\end{tabular}
\end{flushleft}
\end{table*}

Finally, our CXO HRC-S 0.08-10 keV X-ray pulse profile can directly be
compared in absolute phase with our previously derived radio and $\gamma$-ray
profiles \citep{kuip2000}. Firstly, because we generated an ephemeris valid
for all used observation periods and applied a common reference epoch
(see Table \ref{ephemeris}), and secondly, the absolute timing accuracy of CXO is
within $200\mu$s \citep{ten2001} which corresponds to a phase inaccuracy smaller 
than $0.085$ given the pulse period of $2.3$ ms. Fig. {\ref{hepulsprofstack}} 
shows then for the first time the comparison of a \psr X-ray profile in 
absolute phase with the radio and $\gamma$-ray profiles. Within the
uncertainties of each of the measurements the X-ray pulses are aligned with 
two of the three radio pulses at 610 MHz and with the $\gamma$-ray pulses.

\section{Chandra HRC-S X-ray pulse profile vs. EGRET high-energy $\gamma$-ray profile}

The alignment in absolute phase of the very-hard X-ray pulses with the low-significance 
$\gamma$-ray pulses is particularly interesting. The calculated 3.5$\sigma$ significance of the 
$\gamma$-ray profile was derived for {\em unbinned} data giving the statistical 
evidence ($Z^{2}_{4}$ - test) for deviations from a flat distribution {\em anywhere}
in pulsar phase. When we calculate the significance of detection of the two $\gamma$-ray pulses
at the {\em known} absolute phases of the X-ray pulses (e.g. by determining the pulsed excess counts in
the narrow pulse phase windows $0.10$ -- $0.30$ and $0.55$ -- $0.70$, considering the remaining 
phase windows (65\%) as background; we found $54.8\pm 13.7$ pulsed excess counts) then we derive 
for this single trial an increased total significance of 4.0$\sigma$. 

\begin{figure*}
  \hbox{\hspace{1.0cm} \psfig{figure=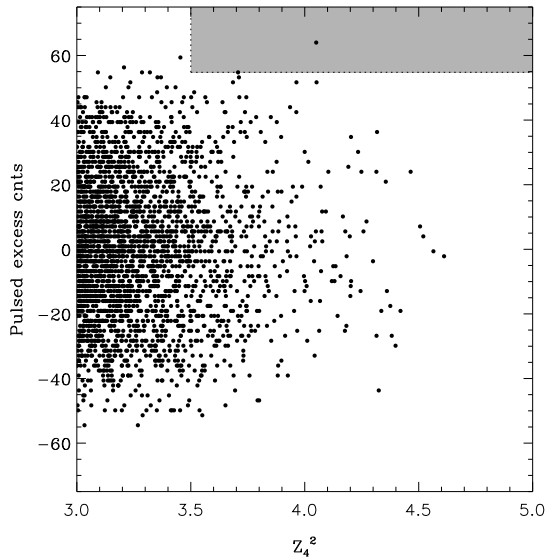,width=7.5cm,height=7.5cm}
  \hspace{1.1cm}{\parbox{80mm}{\vspace{-10cm}
  \caption{Scatter plot of pulsed excess counts vs. $Z_4^2$ for $10^6$ background simulations.
           The hatched band indicates the parameter space of interest, namely, to detect by chance
           a signal with a modulation signifcance above $3.5\sigma$ {\small{\it and}} more than 55 
           pulsed excess counts in the pulse phase windows $0.1$ -- $0.3$ and $0.55$ -- $0.7$ 
           (representative for the 0.1-1 GeV $\gamma$-ray profile). We found just one occurrence in 
           $10^6$ trials, corresponding to a statistical significance of $\sim 4.9\sigma$. 
           \label{egretlcsim} }}}
          }
\end{figure*}

In fact, we performed simulations similar to what we did earlier for PSR B1951$+$32 
\citep{kuip1998b}: We simulated $10^6$ flat phase distributions with the same number of 
counts ($n=308$) as contained in the EGRET 0.1-1 GeV phase histogram shown in Fig. {\ref{hepulsprofstack}}c. 
Then we applied the $Z^{2}_{4}$ - test and verified first that the distribution of the $Z^{2}_{4}$ 
simulations behaved as expected for a random distribution (with $10^6$ trials we could calibrate well up to $\sim 
4.5\sigma$). Next, we used again the two narrow pulse phase windows $0.10$ -- $0.20$ and $0.55$ -- $0.70$ 
to determine the pulsed excess counts.
Fig. \ref{egretlcsim} shows in a two-dimensional distribution all simulations with a 
$Z^{2}_{4}$ - significance above 3$\sigma$ versus the measured pulsed excess counts in the pre-defined
pulse windows. As can be seen, only for one simulation out of $10^6$ a significance above 3.5$\sigma$ 
{\em and} pulsed excess counts above 55 was reached. This corresponds to a statistical significance of $\sim$ 
4.9$\sigma$. Therefore, we take the alignment in absolute phase of the non-thermal X-ray and $\gamma$-ray 
pulses as supporting evidence for our first detection of high-energy $\gamma$-rays from a millisecond pulsar. 

\section{Spectral analysis}

In the spatial and timing analyses we derived the main results from our
Chandra data, which 
contain very limited spectral information. Essentially, we measured the
integral count rate over the entire
0.08-10 keV energy window. Therefore, we cannot add significant new
information to the well-defined pulsed 
spectrum as measured by the BeppoSAX MECS instrument \citep{min2000}.
However, no meaningful information has 
been published sofar on the spectral shape of the DC component (crucial
to decide on the origin and production 
mechanism), while the combined ROSAT HRI, BeppoSAX MECS and Chandra
data can give a first estimate: The ROSAT 
HRI provides us a DC count rate for 0.1-2.4 keV, the MECS for 1.6-4 keV
and 4-10 keV, and Chandra for 0.08-10 keV. 

Assuming a spectral model and absorbing column N$_{\hbox{\rm\scriptsize
H}}$ and using the instrument sensitivity 
curves over the different energy windows, the HRC-S DC count rate can
be used to predict the ROSAT HRI and BeppoSAX 
MECS DC count rates.

For a black body model the measured ROSAT HRI/HRC-S rates demand a
temperature kT $\ltap 0.9$ keV; the measured BeppoSAX MECS/HRC-S rates 
$0.5 \ltap$ kT $\ltap 1.05$ ($1\sigma$ limits). Therefore, a thermal model with a 
temperature $0.5 \ltap$ kT $\ltap 0.9$ (90\% confidence interval: 0.33-1.1 keV)
is consistent with all the measured DC count rates. Assuming a power-law model, the
DC rates of the three instruments are consistent with power-law indices
in the range 1.3-1.85 ($1\sigma$ limits; 90\% confidence interval 0.95-2.45),
significantly softer than for the pulsed emission ($0.61 \pm 0.32$) as
measured by the MECS \citep{min2000}. 
The total statistics of the data are insufficient to discriminate
between the black body and the power-law spectral shapes.

\section{Discussion \& Conclusions}

Our first objective of the Chandra observations was to determine the spatial extent of the X-ray DC component 
and to find a likely explanation for its origin. As is shown in Fig. 1, no evidence for extended emission
near \psr at $\sim 1\arcsec$ scales (diameter) is seen, rejecting the indication for a compact nebula found 
in our analysis of ROSAT HRI data \citep{kuip1998a}. The combined ROSAT, BeppoSAX MECS and Chandra data led 
us to conclude that the DC spectrum is significantly softer than the pulsed spectrum, the latter being clearly 
of non-thermal origin. 
If we adopt the geometry proposed by \citet{nav1995}, a nearly aligned rotator and a small viewing angle, the 
soft DC X-ray component can be explained as thermal emission from the polar cap of the neutron star which stays 
visible to the observer all the time. For old millisecond pulsars the thermal emission can only originate from
reheating of the polar cap area by back-flowing accelerated particles. In fact, for a ms-pulsar the polar-cap half angle $\theta_{pc}$ ($=\arcsin(\sqrt(R_{ns}\cdot 2\pi/p\cdot c))$) extends over a relatively large angle, e.g. $17\fdg 5$ for \psr. This means that the spin-axis is well 
within the (magnetic) polar-cap region, for the measured angle between the spin- and magnetic axes amounts 
$\sim 10\arcdeg$ ($8\arcdeg\pm 11\arcdeg$) \citep{stairs1999}.
Therefore, a small observer viewing angle to the system (angle spin-axis and line of sight) can keep the observer
looking at the heated polar region continuously.

The \citet{nav1995} radio observations had a VLA beam size of
$16\arcsec$, also allowing an interpretation 
of the radio DC component (reported DC fraction $\sim 50\%$) as a
compact nebula up to the size of the VLA
beam. No new observations have been reported with a smaller beam size,
but \citet{stairs1999} and \citet{kuzmin2001} revisited the pulsed
emission. We fitted the total radio 
spectrum of \citet{nav1995} (five data points) with a power-law shape,
index $2.57 \pm 0.07$, and
the new total pulsed radio spectrum (three data points from
\citet{nav1995}, two from \citet{stairs1999} and 
one from \citet{kuzmin2001}), index $2.58\pm 0.15$. The spectra appeared
identical in shape and the updated DC
fraction is $(13 \pm 9)\%$, significantly lower than the earlier
estimate. This leads us to the conclusion 
that the radio DC component is most likely also compact and has the
same magnetospheric origin as the
pulsed emission, which has a very structured and remarkably broad
profile and is indeed practically never 
``off''.

Our second objective was to obtain with Chandra a more significant
X-ray profile than measured sofar, to 
study the profile structure in more detail in the X-ray band below 10
keV, and to have for the first time absolute
timing, allowing multiwavelengths phase comparisons. The profile in
Fig. \ref{hrcspulseprof} with a significance 
of $15.2\sigma$ is indeed much more significant than the MECS profile
($6.8\sigma$). Now we can really see 
that the two pulses have broad wings and that the profile reaches the
DC level only in a very narrow phase interval 
around phase 0.35 (phase extent $\ltap 0.1$). In fact,
Fig.\ref{hepulsprofstack} shows that the radio and the X-ray 
profiles both reach a minimum level for approximately the same {\it
absolute} phase region. Furthermore, Fig.\ref{hepulsprofstack} shows
that the X-ray pulses are aligned in absolute phase with two of the
three radio pulses and
the two $\gamma$-ray pulses within the timing uncertainties of the
different measurements.  

\psr is the first and only millisecond pulsar for which we reported
evidence for detection of high-energy gamma-ray emission up to 1 GeV 
\citep{kuip2000}. We take the alignment in absolute phase of the 
non-thermal X-ray and $\gamma$-ray pulses as important supporting evidence for 
our first detection of high-energy $\gamma$-rays from a millisecond
pulsar. In Sect. 5 we showed that the probability that a random timing signal 
reaches a $3.5\sigma$ modulation significance {\em and} has its 55 pulsed excess 
counts in phase with the two X-ray pulses amounts $\sim 4.9\sigma$.

\citet{wall2000} reported the detection of short-term variability in 
the high-energy $\gamma$-ray emission from 3EG J0222+4253 (3C66A/PSR J0218$+$4232) in a 
systematic search for all 170 unidentified sources in the 3rd EGRET Catalog 
\citep{hart1999}. They noted that if this flaring is due to \psr: ``It would be an 
unusual source in two ways: it would be the only millisecond 
pulsar seen by EGRET and the only pulsar to show strong flaring". We argue, however, 
that in their search the reported evidence for flaring is 
not significant: For each of the 170 unidentified sources \citet{wall2000} 
produced light curves per viewing period (VP) with 2 day flux values each time 
they were in the field-of-view during a VP. For 3EG J0222+4253 (3C66A/PSR J0218$+$4232) 
they found the most significant evidence for variability, namely in VP15 with a 
variability index V=2.6, corresponding to a $\sim 3\sigma$ significance for a 
random detection in a {\em single} trial assuming Gaussian statistics (although Poisson 
statistics apply). This variability was due to one single high 2 day flux value. 
However, this source was viewed in 4 observations (no indication for variability 
in the other VPs) which makes the probability ($4\times 0.002512$) 1\% or 2.58$\sigma$ to 
find V=2.6 in 1 out of 4 observations. Furthermore, \citet{wall2000} analysed 144 VPs 
with duration larger than 3 days producing for the 170 unidentified sources 
few-hundred light curves, making the probability to find once V=2.6 (or one single high 2 day flux value)
few times unity! 
They correctly noted that their Monte Carlo probabilities give misleading values 
(too optimistic) as the used averages having no uncertainties assigned. Also in the 
latter case the very large number of ``trials'' has been ignored. Since there was 
no {\em a priori} reason to select 3EG J0222+4253 for a single trial, we donot regard 
the indication for short-term variability of this source significant.

The X-ray results on the DC emission and the radio results on the DC and pulsed emission
suggest for \psr an emission scenario of a nearly aligned rotator and a small viewing angle. 
The latter angle can still be as large as $\sim 20\arcdeg$ to explain the X-ray DC emission
if this originates from a polar-cap heated by particle bombardment. 
For explaining the pulse profile with two hard-spectrum X-ray/$\gamma$-ray peaks it is more critical to
know this viewing angle to get a handle on the geometry. 
For both competing classes of models, polar cap models (PC; e.g. \citet{daugherty94,daugherty96}) 
and outer gap models (OG; see e.g. \citet{cheng86a,cheng86b,ho89}) production of hard X-ray/$\gamma$-ray 
emission in the magnetospheres of millisecond pulsars has been predicted (e.g. \citet{bhattacharya91,sturner94}). 
Recently, millisecond pulsars among them \psr were considered for PC models by \citet{dyks99,bulik99,bulik00,zhang00}. 
For \psr they did not succeed in reproducing the measured high-energy spectral shape. More recently, \citet{dyks02} 
succeeded in reproducing the high-energy spectrum of PSR J0218$+$4232, but required non-orthodox assumptions about the electron energy distribution or emission altitude as well as off-beam viewing geometry. \citet{wozna02}
reproduced the double-peak profile with the measured phase separation for a small inclination angle 
of $8\arcdeg$ and a viewing angle of $29\arcdeg$ (angles approximately consistent with our findings). 
However, for these angles the spectrum reached its maximum luminosity for too high energies around 
100 GeV. The fact that in the case of PC models a double-peak hard X-ray/$\gamma$-ray profile can 
be modelled for a nearly aligned rotator with a $\sim 20\arcdeg$ viewing angle follows already 
from \citet{daugherty96}, who showed that the $\gamma$-ray beam produced at the polar cap rim 
is relatively wide, amounting $\sim 26 \arcdeg$ ($1.5 \times \theta_{pc}$) for PSR J0218$+$4232. 
For emission produced at higher altitudes in the magnetosphere (as is proposed in the more recent 
versions of PC models), the beam will become even wider. In all these discussions the real viewing 
angle in combination with the inclination angle determines whether the measured double peak profile can be 
obtained. Also for OG models a double peak profile is in principle possible, depending on 
the actual viewing angle and which part of the outer gap (in altitude) is visible to the observer. For OG models the 
hard X-ray/$\gamma$-ray cone might become too broad for the case of \psr as a nearly aligned rotator if the hard X-ray/$\gamma$-ray production takes place only close to the light cylinder. However, \citet{hirotani02} 
argued recently that the $\gamma$-ray production is not limited to regions in the outer gap above the 
null-charge surface, but can also originate from altitudes closer to the neutron star. More detailed
model calculations are required to solve the present uncertainties in the overall interpretation of the data.

This underlines the importance of a new attempt to determine the geometry from radio polarization data.
Then detailed model calculations starting from this geometry can attempt to reproduce all the remarkable 
timing and spectral results obtained for this millisecond pulsar.  

\acknowledgments

\clearpage 
\end{document}